\documentclass[10pt, conference, letterpaper]{IEEEtran}
\IEEEoverridecommandlockouts

\usepackage{amsmath, amssymb, amsfonts, mathtools} 
\usepackage{graphicx} 

\usepackage{comment} 
\usepackage{subfigure}
\usepackage{algorithm}
\usepackage{algorithmicx}
\usepackage{algpseudocode}
\usepackage{cuted}
\usepackage[table]{xcolor}
\usepackage{colortbl}
\usepackage{array, booktabs, multirow, makecell} 
\usepackage{caption} 
\usepackage{float} 
\usepackage{placeins} 
\usepackage{threeparttable} 
\usepackage{longtable} 
\usepackage{adjustbox} 

\usepackage{tikz} 
\usepackage{tkz-euclide} 
\usetikzlibrary{shapes.geometric, 3d} 
\usepackage{pdflscape} 
\usepackage{enumitem} 
\usepackage{etoolbox} 

\usepackage{cite}
\def\BibTeX{{\rm B\kern-.05em{\sc i\kern-.025em b}\kern-.08em
    T\kern-.1667em\lower.7ex\hbox{E}\kern-.125emX}}

\graphicspath{{./Images/}} 

\usepackage{pifont} 
\definecolor{myred}{RGB}{226,13,118}
\definecolor{myblue}{RGB}{0,112,196}
\definecolor{mynaiveblue}{RGB}{117,232,255}
\usepackage{dirtytalk}
\usepackage{soul}
\begin{document}

\title{Discrete Mode Decomposition Meets Shapley Value: Robust Signal Prediction in Tactile Internet}

\author{\IEEEauthorblockN{Mohammad Ali Vahedifar and Qi Zhang}
\IEEEauthorblockA{DIGIT and Department of Electrical and Computer Engineering, Aarhus University, Denmark}
\thanks{This research was supported by the TOAST project, funded by the European Union’s Horizon Europe research and innovation program under the Marie Skłodowska-Curie Actions Doctoral Network (Grant Agreement No. 101073465), the Danish Council for Independent Research project eTouch (Grant No. 1127- 00339B), and NordForsk Nordic University Cooperation on Edge Intelligence (Grant No. 168043). Authors' e-mails: \{av, qz\}@ece.au.dk.}
}

\maketitle
 \begin{abstract}

Tactile Internet (TI) requires ultra-low latency and high reliability to ensure stability and transparency in touch-enabled teleoperation. However, variable delays and packet loss present significant challenges to maintaining immersive haptic communication. To address this, we propose a predictive framework that integrates Discrete Mode Decomposition (DMD) with Shapley Mode Value (SMV) for accurate and timely haptic signal prediction. DMD decomposes haptic signals into interpretable intrinsic modes, while SMV evaluates each mode’s contribution to prediction accuracy, which is well-aligned with the goal-oriented semantic communication. Integrating SMV with DMD further accelerates inference, enabling efficient communication and smooth teleoperation even under adverse network conditions.

Extensive experiments show that DMD+SMV, combined with a Transformer architecture, outperforms baseline methods significantly. It achieves 98.9\% accuracy for 1-sample prediction and 92.5\% for 100-sample prediction, as well as extremely low inference latency: 0.056 ms and 2 ms, respectively. These results demonstrate that the proposed framework has strong potential to ease the stringent latency and reliability requirements of TI without compromising performance, highlighting its feasibility for real-world deployment in TI systems.
\end{abstract}


\IEEEpeerreviewmaketitle

\section{Introduction}

Tactile Internet (TI) allows users to interact with physical or virtual objects by enabling the communication of touch sensations over large distances, which is expected to support diverse emerging mission-critical applications, such as remote surgery and robotic manipulation~\cite{zhang20185genabledtactilerobotic}. To ensure immersive operation of these applications and keep
the communication delay as small as possible, TI systems must achieve ultra-low end-to-end latency (around 1~ms) for haptic packets due to haptic sensor readings, have a sampling rate of 1~kHz or even higher~\cite{kokkinis2025deepreinforcementlearningbasedvideohaptic}. However, achieving such stringent latency requirements remains a significant challenge in wireless communication networks due to the time-varying channel and network conditions~\cite{ GTvahedifar2025shapleyfeaturesrobustsignal}. Delay in haptic packets or packet loss can cause instability and compromise transparency in haptic control, for example, resulting in deep penetration and cybersickness, which directly impacts task precision and safety~\cite{promwongsa}.

In scenarios where consistent real-time human control is temporarily not available, it is necessary to incorporate semi-autonomous systems that can learn and predict the actions of both users and robots within short time windows~\cite{Antonakoglou}. These systems can assist or temporarily replace human operators, ensuring continuity and stability in control even under harsh network conditions.
While edge intelligence is a promising solution for signal prediction, existing solutions typically rely on historical signals ~\cite{LeFo, kokkinis2025deepreinforcementlearningbasedvideohaptic}. This approach faces significant challenges when the underlying signal distribution changes rapidly.

This limitation motivates us to propose Discrete Mode Decomposition~(DMD) to decompose discrete signals in TI into fundamental components. This method enables us to isolate and extract meaningful patterns or ``modes" from a complex signal. To evaluate the importance of each extracted mode, we introduce the Shapley Mode Value (SMV), inspired by the Shapley value concept from cooperative Game Theory~(GT). SMV quantitatively assesses the contribution of each mode to the overall prediction task, allowing us to identify the most informative components of the signal. 

Our key contribution is the proposal of DMD and a combination of DMD and SMV to effectively and accurately predict haptic signals to mitigate the impact of communication delays or packet loss caused by temporary communication disruptions. This approach is particularly beneficial in scenarios where communication channels are unreliable or when semi-autonomous system behavior is necessary in the absence of timely input. With this intelligent function at both local and remote sides, we can significantly relax communication latency and reliability constraints in TI, which can be leveraged to increase network capacity, thereby accommodating more users in the system. Notably, our DMD and DMD+SMV methods are not limited to haptic signals only, which can be used for other discrete signal decompositions.

\section{Related Works}\label{RW}
The integration of predictive models into teleoperation systems is a key focus of current research; these models offer promising avenues for more intuitive and practical remote control. For instance, Xu et al.~\cite{xu2020error} proposed Remedy LSTM for haptic packet prediction, achieving higher accuracy than linear estimators but relying on a normality assumption and lacking explicit support for missing data. Salvato et al.~\cite{salvato2022predicting} introduced a self-attention model to predict hand–object contact timing, reducing actuator-induced delays; however, it was limited to discrete events and did not address continuous force trajectory prediction. Kizilkaya et al.~\cite{kizilkaya2023task} used GAN-generated data to augment predictive models, enabling analysis of factors such as JND and prediction error, though the fidelity of synthetic data for rare or complex haptic patterns remains a concern. Kokkinis et al.~\cite{kokkinis2025delayboundrelaxationdeep} introduced a CNN-LSTM-Transformer-based architecture combining remote force measurements with operator motion data for force feedback prediction.  Despite its advancements, a notable limitation is the model's focus on predicting force feedback from the robot side. Vahedifar et al.~\cite{LeFo} leveraged game-theoretic models, including Stackelberg and MinMax formulations, to predict haptic signals bidirectionally. However, the method’s exhaustive prediction of all features may be inefficient. A more selective approach based on feature importance could offer a better trade-off between accuracy and computational demand.

The concept of the Shapley value was first introduced in a classical paper~\cite{shapley:book1952}. Recently, Shapley values have become a popular tool for determining the importance of each data point~\cite{Data-Shapleyghorbani19c} or feature~\cite{Lundberg2017} to the model’s output. This is especially useful when working with black-box models.  Another relevant area is signal decomposition, where advanced adaptive techniques enable the separation of complex waveforms into constituent components. Among these approaches, Variational Mode Decomposition (VMD)~\cite{DragomiretskiyVMD} has emerged as a robust framework that employs constrained variational optimization to extract amplitude-modulated and frequency-modulated components while minimizing mode bandwidths for continuous signals. However, a key limitation is that VMD needs to pass the number of modes before extracting them. Building upon VMD, extensions such as Variational Mode Extraction (VME)~\cite{NazariVME}, and Successive Variational Mode Decomposition (SVMD)~\cite{SVMDNAZARI2020107610} address parameter sensitivity challenges through iterative mode extraction and residual analysis. The key limitation of VME and SVMD is that both algorithms introduce unnecessary hyperparameters into their optimization problems, which create sensitivity and additional computational overhead when determining the corresponding penalty parameters. Our proposed DMD and DMD+SMV address the gap identified in the previous works.
\section{System Model}\label{Section: system model}
\begin{figure}[tbp]
    \centering
    \includegraphics[width=1.1\linewidth]{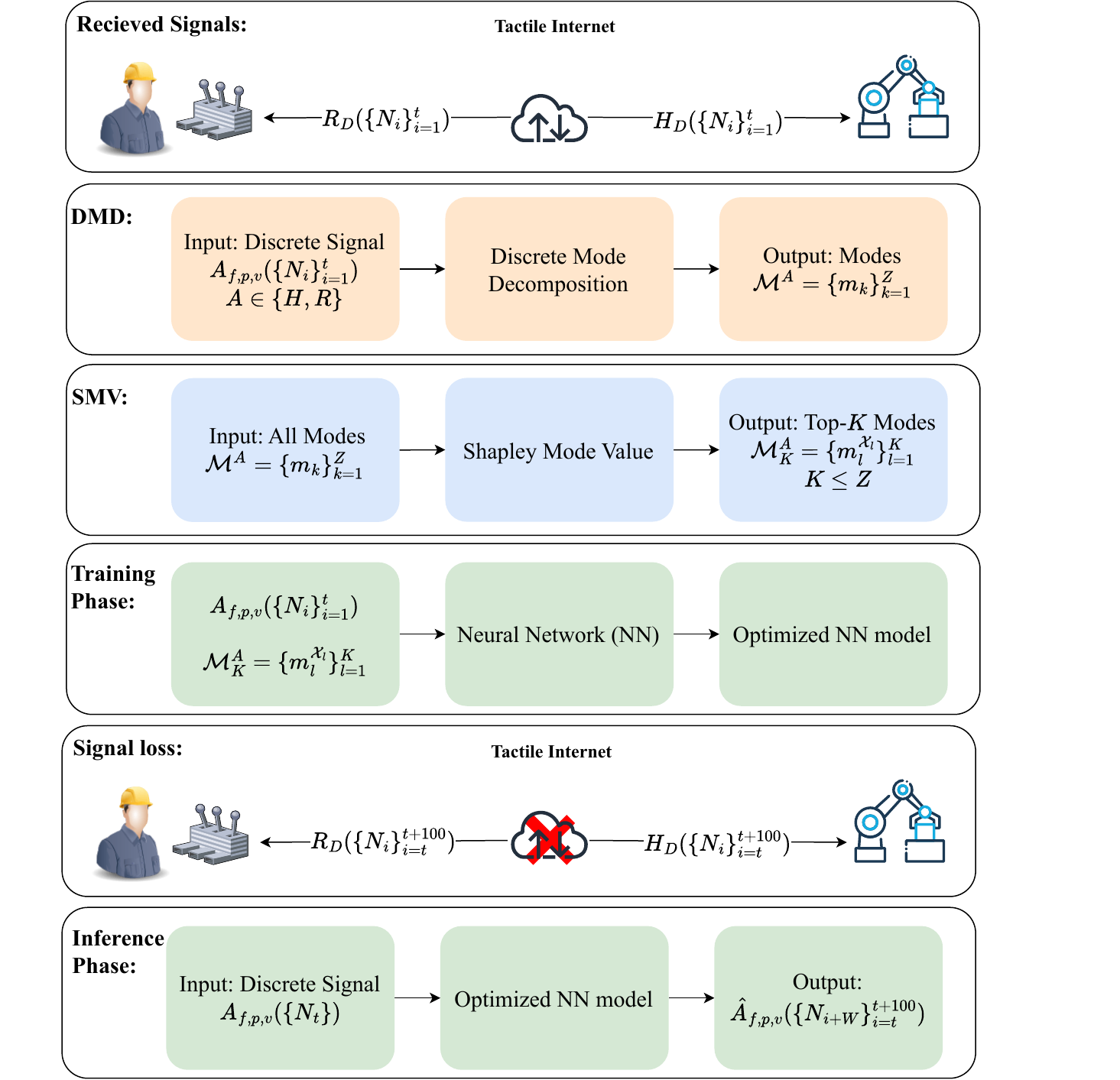}
    \caption{Overview of three-stage mode-centric predictive framework for TI.}
    \label{fig: DMD conceptual fig}
\end{figure}
As illustrated in Fig.~\ref{fig: DMD conceptual fig}, we consider a teleoperation system in which a human~($H$) operator controls a remote robot~($R$) through bidirectional signal exchange. We assume that the input signals denoted as \( A_{f,p,v}(\{N_i\}^t_{i=1} )\), where \( A \in \{H, R\} \) corresponds to the human and robot sides, ${f,p,v}$ denotes force, position velocity, respectively. The input undergoes decomposition into fundamental modes via DMD, resulting in a set of modes \( \mathcal{M}^A=\{m_k\}_{k=1}^Z \), where $Z$ is unknown a priori number of modes. The relative importance of top-$K$ modes is then quantified using SMV, producing a ranked subset \( \mathcal{M}_K^A=\{{m_l^{\mathcal{X}_l}}\}_{l=1}^K \), where \( K \leq Z \) and sorted by the descending significance of \(\mathcal{X}_l \). The sorted modes, together with the discrete input signal information, explicitly guide the Neural Network (NN) in learning the underlying mapping function. During the inference phase, upon receiving the initial input, namely, $A_{f,p,v}(\{N_{t}\})$, the model predicts the signal values for the subsequent time horizon, depending on a defined window size of up to 100 samples in this paper. This yields predicted outputs $\hat{A}_{f,p,v}(\{N_{i+W}\}^{t+100}_{i=t})$. Once a predefined number of samples has been processed, the modes are updated.
\section{Discrete Mode Decomposition}\label{DMD}
Mode decomposition is a technique to decompose a signal into its constituent intrinsic modes. In the following, we briefly review several fundamental concepts and tools from signal processing that will serve as the key building blocks of our DMD model. We first present the definition of the intrinsic mode function, Wiener filtering, and the Hilbert transform, then we propose our DMD.
\subsection{Intrinsic Mode Functions}
For a discrete time signal, Intrinsic Mode Functions (IMFs) are represented as amplitude-modulated frequency-modulated signals of the form:
\begin{equation}
x[n] = A[n] \cos(\phi[n]),
\end{equation}
where the phase \( \phi[n] \) is a monotonically increasing sequence, i.e., \( \phi[n+1] - \phi[n] \geq 0 \). The amplitude envelope satisfies \( A[n] \geq 0 \). The instantaneous frequency is defined as \( \omega[n] = \phi[n+1] - \phi[n] \). 
Note, both \( A[n] \) and \( \omega[n] \) vary slowly relative to the phase \( \phi[n] \). In other words, over a sufficiently large discrete interval \( [n - \Delta n, n + \Delta n] \), where \( \Delta n \approx \frac{2\pi}{\omega[n]} \), the IMF can be treated as a locally harmonic signal with smoothly varying amplitude and frequency~\cite{oppenheim1999discrete}.

\subsection{Discrete Wiener Filtering}\label{Discrete Wiener Filtering}
Consider the observed discrete-time signal $y[n]$, a version of the original signal $x[n]$ corrupted by additive zero-mean white Gaussian noise $\eta[n]$:
\begin{equation}
    y[n] = x[n] + \eta[n], \quad \eta[n] \sim \mathcal{N}(0, \alpha).
\end{equation}
The denoising problem can be formulated as the following discrete Tikhonov regularization problem~\cite{Tikhonov:1963}:

\begin{equation}
    \min_x \left\{\big\|x[n] - y[n]\big\|_2^2 + \alpha  \big\|\nabla x[n]\big\|_2^2 \right\}.
\end{equation}
The solution is obtainable by the Euler-Lagrange equations in the discrete Fourier domain:

\begin{align}
    &X(\omega) = \frac{Y(\omega)}{1 + \alpha |\omega|^2},\\
    &X(\omega) = \sum_{n=-\infty}^{\infty} x[n] \, e^{-j\omega n}, \quad \omega \in [-\pi, \pi].
\end{align}
Here, $\alpha$ represents the variance of white noise. Indeed, the solution corresponds to convolution with a Wiener filter, and the original signal $x[n]$ has a low-pass power spectrum prior $1/|\omega|^2$~\cite{Hahn1996}.

\subsection{Discrete Hilbert Transform}

The N-point Hilbert transform is characterized by its Discrete Fourier Transform (DFT) coefficients and impulse response~\cite{wiener2013extrapolation}:
\begin{equation}\label{Hilbert_frequency}
    H[k] = 
    \begin{cases}
        -j & \text{for } 1 \leq k < N/2 \\
        0 & \text{for } k = 0, N/2 \\
        j & \text{for } N/2 < k \leq N-1
    \end{cases}
\end{equation}
\begin{equation}
h[n] = 
\begin{cases} 
\frac{2\sin^2(\pi n / 2)}{\pi n} = \frac{1 - (-1)^n}{\pi n}, & n \neq 0 \\
0, & n = 0
\end{cases}
\end{equation}
The analytic signal $q[n]$ associated with a real discrete signal $x[n]$ is constructed as:
\begin{equation}
    q[n] = x[n] + jH\{x[n]\}.
\end{equation}
\subsection{Discrete Mode Decomposition}
Let $x[n]$ be the discrete-time input signal of length $N$. The goal of DMD is to decompose a real-valued input signal \( x[n] \) into a set of discrete mode, \( \mathcal{M}^A=\{m_k\}_{k=1}^Z \), each centered around a specific frequency, \( \mathcal{W}^A=\{\omega_k\}_{k=1}^Z \). To mathematically represent the method, we assume that the input signal $x[n]$ is decomposed into three signals: 1)~The $Z$-th mode ($m_Z[n]$). 2)~The summation of previously extracted modes ($\sum^{Z-1}_{k=1} m_k[n]$). 3)~The unprocessed part of the signal ($x_u[n]$). Subsequently, the original signal in the discrete time domain is:
\begin{equation}
x[n] = \sum_{k=1}^{Z} m_k[n] + x_u[n]= m_Z[n] + \sum_{k=1}^{Z-1} m_k[n] + x_u[n].
\end{equation}
This iterative extraction continues until the reconstruction error (the difference between the original signal and the sum of the extracted modes) below a specified threshold. We now present our proposed decomposition method based on five key criteria.

\noindent \textbf{1.~Spectral Compactness of $Z$-th mode:} Each mode should be compact around its center frequency. Consequently, the $Z$-th mode minimizes the following criterion:
\begin{equation}\label{T_1}
 \hspace{-2mm}T_1=\bigg\| \partial_n \Big[\Big(\delta[n] + j (\frac{1 - (-1)^n}{\pi n})\Big)* m_Z[n]\Big) e^{-j \omega_Z n} \Big] \bigg\|_2^2.
\end{equation}
Here, the Hilbert transform is applied to $m_Z$ to obtain a one-sided frequency spectrum, then shifts the spectrum to center around zero frequency. This is done by multiplying the mode by $e^{-j\omega_Z n}$. Then the bandwidth is quantified by the Gaussian smoothness of the demodulated signal, computed as the squared $L^2$-norm of its centered difference discrete derivative. A smaller gradient norm indicates a smoother (more compact) frequency spectrum, implying a narrower bandwidth.

\noindent \textbf{2.~Minimum Overlap with Previously Extracted Modes:} This encourages spectral separation between the current mode and all previously extracted modes. Consequently,  $m_Z(n)$ should have less energy at frequencies around the center frequencies of the previously obtained modes:
\begin{align}
&T_2 = \sum_{k=1}^{Z-1} \Big\| \beta_k[n] * m_Z[n] \Big\|_2^2,\label{T_2}\\
&\beta_k(\omega) = \frac{1}{\alpha(\omega - \omega_k)^2 + \epsilon_1},\quad k=1,2,\cdots,Z-1,
\end{align}
where $\beta_k(\omega)$ is the frequency response of the filter in Eq.~\ref{T_2}, $\epsilon_1$ is a small regularization constant ($\epsilon_1 \ll 1$), and $\alpha$ represents the variance of white noise as in subsection~\ref{Discrete Wiener Filtering}.

\noindent \textbf{3.~Minimum Spectral Overlap with Unprocessed Signal:} The energy of the unprocessed signal $x_u[n]$ should be minimized at frequencies where the $Z$-th mode $m_Z[n]$ has significant components. This constraint is implemented using a discrete filter $\beta_Z[n]$:
\begin{align}
&T_3 = \Big\| \beta_Z[n]  *x_u[n] \Big\|_2^2,\label{T_3}\\
&\beta_Z[\omega] = \frac{1}{\alpha(\omega - \omega_Z)^2 + \epsilon_2},
\end{align}
where $\beta_Z(\omega)$ is the frequency response of the filter in Eq.~\ref{T_3}, and $\epsilon_2$ is a small regularization constant ($\epsilon_2 \ll 1$).

\noindent \textbf{4.~Reconstruction Constraint:} This constraint is to guarantee complete reconstruction of $x[n]$ from $Z$ modes of the signal:
\begin{equation}
x[n] =  \sum_{k=1}^{Z} m_k[n] .
\end{equation}
\noindent \textbf{5.~Bounded Unprocessed Energy:} To ensure the decomposition is meaningful and the unprocessed part is negligible, the energy of the unprocessed signal must not exceed the energy of the least significant extracted mode.
\begin{equation}
\big\|x_u[n]\big\|_2^2 \leq \big\|m_{\text{min}}[n]\big\|_2^2, \quad  m_{\text{min}} = \arg\min_{m \in \mathcal{M}} \|m\|_2.
\end{equation}
\subsection{Overall Optimization Problem}
Based on the criteria outlined above, we formulate the primary optimization problem.
\begin{equation}
\begin{split}\label{Eq: original optimization problem}
 \hspace{-5mm}\min_{m_Z, \omega_Z} \quad 
 &T_1 + T_2 + T_3\\
s.t \quad &x[n] =  \sum_{k=1}^{Z} m_k[n] + x_u[n],\\
& \big\|x_u[n]\big\|_2^2 \leq \big\|m_{\text{min}}[n]\big\|_2^2, \quad  m_{\text{min}} = \arg\min_{m \in \mathcal{M}} \|m\|_2.
\end{split}
\end{equation}
\noindent For the next step, we utilize the augmented Lagrangian.
\begin{equation}\label{Eq: final Lagrange Optimization problem}
\begin{aligned}
\mathcal{L}_{\text{aug}}\big(\mu,\rho\big) &= \big\|\partial_n \big[A_n * m_Z \big] e^{-j\omega_Z n}\big\|_2^2 + \sum_{k=1}^{Z-1} \big\| \beta_k * m_Z \big\|_2^2 \\
&+ \big\| \beta_Z * x_u \big\|_2^2 + \frac{\rho}{2} \big\|x - \sum_{k=1}^{Z} m_k - x_u + \theta\big\|_2^2 \\
&+ \mu\big(\|x_u\|_2^2 - \|m_{\text{min}}\|_2^2\big) - \frac{\rho}{2} \big\|\theta\big\|_2^2.
\end{aligned}
\end{equation}
\noindent where $A_n = \delta[n] + j (\frac{1 - (-1)^n}{\pi n})$ . All variables are functions of $n$ unless otherwise noted. Note, $\mu$ is the scalar Lagrangian multiplier and $\frac{\rho}{2}$ is the penalty parameter control, which, by applying a variable change $\lambda[n]=\rho[n]\theta[n]$, becomes $\lambda[n]$ Lagrange multiplier, where $\theta[n]$ is the scaled dual variable. Augmented Lagrangian combines the benefits of quadratic penalties (for convergence at finite weights) and Lagrangian multipliers (for strict constraint enforcement). The solution to the original optimization problem Eq.~\ref{Eq: original optimization problem} is now found as the saddle point of the augmented Lagrangian Eq.~\ref{Eq: final Lagrange Optimization problem}. For the next step, we transform the Eq.~\ref{Eq: final Lagrange Optimization problem} to the Frequency domain.
Note we used the Parseval Theorem, centered difference discrete derivative, $|(1+\text{sgn}(\omega))|^2=4$, and variable change for the first term $\omega \leftarrow \omega-\omega_Z$.
\begin{equation}\label{Eq: Lagrange Frequency Domain}
\begin{aligned}
&\mathcal{L}_{\text{aug}}\big(\mu,\rho(\omega)\big) =\frac{2}{\pi}\int_0^\pi  \sin^2(\omega-\omega_Z) \Big|M_Z(\omega )\Big|^2 \, d\omega \\&+\sum_{k=1}^{Z-1} \int_0^\pi \big\|\beta_k(\omega) M_Z(\omega)\big\|_2^2  d\omega + \int_0^\pi \big\|\beta_Z(\omega) X_u(\omega)\big\|_2^2  d\omega
\\&+ \frac{\rho(\omega)}{2}\Big\| X(\omega) -  \sum_{k=1}^{Z} M_k(\omega)-X_u(\omega) + \Theta(\omega) \Big\|_2^2\\&+\mu \big(\big\|X_u(\omega)\big\|_2^2 - \big\|M_{\text{min}}(\omega)\big\|_2^2 \big) - \frac{\rho(\omega)}{2}\big\|\Theta(\omega)\big\|_2^2 ~.
\end{aligned}
\end{equation}
The optimization problem in Eq.~\ref{Eq: Lagrange Frequency Domain} is solved by an iterative approach called the Alternating Direction Method of Multipliers (ADMM)~\cite{Bertsekas1886529043}. This method breaks down the complex problem into simpler sub-problems that are easier to solve individually. In the following, we will explain step-by-step how each of these sub-problems can be solved effectively. 
\textbf{Update $M_Z(\omega)$:} We minimize Eq.~\ref{Eq: Lagrange Frequency Domain} with respect to $M_Z(\omega)$, keeping all other variables fixed. Extract the relevant terms:

\begin{equation}
\begin{aligned}
 &\arg\min_{M_Z}\Big\{\frac{2}{\pi}\int_0^\pi \sin^2(\omega-\omega_Z)\Big|M_Z(\omega )\Big|^2 \, d\omega \\
  &+ \sum_{k=1}^{Z-1} \int_0^\pi \big|\beta_k(\omega) M_Z(\omega)\big|^2  d\omega \\
 &+ \frac{\rho(\omega)}{2} \big| X(\omega) - M_Z(\omega) - \sum_{k=1}^{Z-1} M_k(\omega) -X_u(\omega)+ \Theta(\omega) \big|^2
\Big\}.
\end{aligned}
\end{equation}
Let the auxiliary variable be:
\begin{equation}
Q(\omega) = X(\omega) - \sum_{k=1}^{Z-1} M_k(\omega) -X_u(\omega)+ \Theta(\omega).
\end{equation}
Now, letting the first variation vanish for the positive frequencies leads to the closed-form solution in the $n+1$-th iteration:
\begin{equation}\label{Eq.U_L}
\begin{split}
M_Z^{n+1}(\omega) = \frac{\frac{\rho(\omega)}{2} Q(\omega)}{\frac{\rho(\omega)}{2} +\sum_{k=1}^{Z-1} \big| \beta_k(\omega)\big|^2 + \frac{2}{\pi} \sin^2(\omega-\omega_Z)}.
\end{split}
\end{equation}
\noindent\textbf{Update $\omega_Z$:}
We minimize Eq.~\ref{Eq: Lagrange Frequency Domain} concerning $\omega_Z$ and substitute~\ref {T_3}, keeping all other variables fixed. Extract the relevant terms:
\begin{equation}
\begin{split}
  \arg\min_{\omega_Z}\Big\{\frac{2}{\pi}\int_0^\pi  \sin^2(\omega-\omega_Z) \Big|M^{n+1}_Z(\omega )\Big|^2  d\omega\\
+  \int_0^\pi \big|\frac{X_u(\omega)}{\alpha(\omega - \omega_Z)^2 + \epsilon_2} \big|^2  d\omega \Big\}.
\end{split}
\end{equation}
Ignoring the second term as it is significantly smaller than the first term (in practice, $\alpha\gg1$ to enforce strict spectral compactness), we will have:
\begin{equation}
\arg\min_{\omega_Z}\Big\{\frac{2}{\pi}\int_0^\pi  \sin^2(\omega-\omega_Z) \Big|M^{n+1}_Z(\omega )\Big|^2 \, d\omega \Big\}.
\end{equation}
Considering Taylor expansion for $\sin(\omega-\omega_Z)$. This quadratic problem is solved as:
\begin{equation}\label{Eq.W_L}
\omega_Z^{n+1} = \frac{ \int_{0}^{\pi} \omega | M^{n+1}_Z(\omega)|^2 \, d\omega }{ \int_{0}^{\pi} | M^{n+1}_Z(\omega)|^2 \, d\omega }.
\end{equation}
\textbf{Update $X_u(\omega)$:} We minimize Eq.~\ref{Eq: Lagrange Frequency Domain} with respect to \(X_u(\omega) \), keeping all other variables fixed. Extract the relevant terms:
\begin{equation}
    \begin{split}
         \arg \hspace{-1mm}\min_{X_u(\omega)} \Big\{\int_0^\pi \big\|\beta_Z(\omega) X_u(\omega)\big\|_2^2 \, d\omega + \mu\big\|X_u(\omega)\big\|_2^2\\
         + \frac{\rho(\omega)}{2}\Big\| X(\omega) -  \sum_{k=1}^{Z} M_k(\omega)-X_u(\omega) + \Theta(\omega) \Big\|_2^2\Big\}.
    \end{split}
\end{equation}
Let the auxiliary variable be:
\begin{equation}
 \widetilde{Q}(\omega)=X(\omega) -  \sum_{k=1}^{Z} M_k(\omega) + \Theta(\omega).
\end{equation}
The optimal magnitude for $X_u(\omega)$ is:
\begin{equation}\label{Eq.F_u}
X_u^{n+1}(\omega) = \frac{\rho(\omega)\widetilde{Q}(\omega)}{2|\beta_Z(\omega)|^2+2\mu+\rho(\omega)}.
\end{equation}

\noindent\textbf{Update $\Theta(\omega)$:} This yields to update of equality constraint $\lambda[n]=\rho[n]\theta[n]$ which is updates through dual ascent method~\cite{Bertsekas1886529043}:
\begin{equation}\label{Eq.P_k}
\begin{split}
\Theta^{n+1}(\omega) =\Theta^{n}(\omega) + \tau_1\hspace{-1mm}\left(\hspace{-1mm} X(\omega)\hspace{-1mm}-\hspace{-1mm}\sum_{k=1}^{Z} M^{n+1}_k(\omega)\hspace{-1mm}-\hspace{-1mm}X_u^{n+1}(\omega) \right).
\end{split}
\end{equation}
\textbf{Update $\mu$:} This is update of inequality constraint.
\begin{equation}\label{Eq.M_k}
\mu^{n+1} \hspace{-1mm}= \max\hspace{-1mm}\left(\hspace{-1mm}0, \mu^{n} \hspace{-1mm}+\hspace{-1mm} \tau_2 \int_0^\pi\hspace{-1mm}\left(\|X^{n+1}_u(\omega)\|^2_2 \hspace{-1mm}-\hspace{-1mm} \|M_{\text{min}}(\omega)\|^2_2 \right) d\omega \right).
\end{equation}
Accordingly, Algorithm~\ref{algorithm DMD} summarizes the steps of the DMD.
\begin{algorithm}[t]
\caption{Discrete Mode Decomposition (DMD)}\label{algorithm DMD}
\begin{algorithmic}
\State \textbf{Input:} Discrete signal $x[n]$, noise variance $\alpha$
\State \textbf{Initialize:} Parameters $\epsilon_1, \epsilon_2$, $\tau_1\gets1$, $\tau_2\gets1$, $\kappa_1$, $\kappa_2$
\Repeat
\State \textbf{Initialize:} $m^1_Z,\omega^1_Z,\mu^1$, $n\gets0$, $Z\gets0$

\State $Z\gets Z+1$
\Repeat
\State $n \gets n+1$        
\State Update $M_Z(\omega)$ with Eq.~\ref{Eq.U_L}
\State Update $\omega_Z$ with Eq.~\ref{Eq.W_L}
\State Update scaled dual variable with Eq.~\ref{Eq.P_k}
\State Update inequality constraint with Eq.~\ref{Eq.M_k} 
\Until{$\frac{\|M_Z^{n+1} - M_Z^{n}\|^2_2}{\|M_Z^{n}\|^2_2}\leq \kappa_1$}
\Until{$\frac{|\alpha-\frac{1}{Z}\|x-\Sigma^Z_{k=1}m_k\|^2_2}{\alpha}\leq \kappa_2$}
\State \textbf{Output:} $\mathcal{M^A,W^A}$
\end{algorithmic}
\end{algorithm}

\section{Shapley Mode Value}\label{SMV}
A critical challenge in ML-based signal prediction is quantifying the contribution of individual features or extracted modes to model performance. Traditional approaches often favor simpler, interpretable models (e.g., linear regression) at the expense of predictive accuracy, as they provide clearer insights into decision-making. However, with the growing complexity of haptic datasets, deep learning models can capture nuanced patterns that simpler methods miss, while their opacity hinders trust, optimization, and real-world deployment.

This motivates us to propose a method for quantifying precisely the contribution of each mode to the overall learning process in DMD for signal prediction. We propose \textbf{Shapley Mode Value (SMV)}, inspired by the Shapley value. This helps resolve the accuracy versus interpretability dilemma by providing a systematic way to understand what determines a complex model's performance, even when the model itself is complex to interpret directly. Our approach achieves two key objectives: \textbf{1.~Accelerating Inference:} By identifying and retaining only task-relevant features, we reduce computational overhead without sacrificing accuracy. \textbf{2.~Enhancing Accuracy:} Shapley values prioritize meaningful signal modes, discarding redundant or noisy modes.

This method, well aligned with the goal-oriented communication paradigm, not only improves interpretability but also optimizes resource utilization, a crucial advantage for latency-sensitive applications like haptic feedback in 5G/6G networks. Furthermore, it bridges the gap between model performance and actionable insights, ensuring explainability in predictions.

\subsection{Shapley Mode Value (SMV)}

Let $\mathcal{D} = \{(m_k, \omega_k)\}_{k=1}^Z$ be our fixed training set containing our modes and their central frequency. Let $\mathcal{Z}$ denote the DMD algorithm, and $\mathcal{S} \subseteq \mathcal{D}$. The performance score $V$ is a black-box oracle that evaluates a predictor and returns a score. We denote $V(\mathcal{S}, \mathcal{Z})$ as the performance score of the predictor trained on the mode set $\mathcal{S}$. Our goal is to compute a mode value $\mathcal{X}_k(\mathcal{D}, \mathcal{Z}, V) \in \mathbb{R}$, to quantify the value of the $k$-th mode. For simplicity, we write $\mathcal{X}_k$. We define that $\mathcal{X}_k$ should satisfy the following properties:

\noindent\textbf{1. Transferability:} \label{axiom 1} The total value is distributed among all modes:
    \begin{equation}
    \sum_{k \in \mathcal{D}} \mathcal{X}_k = V(\mathcal{D}),
    \end{equation}
    i.e., the sum of individual values equals the total value.
    
\noindent\textbf{2. Monotonicity:} This axiom satisfies three axioms simultaneously: \textbf{a. Null contribution:} If adding a specific feature does not improve performance, no matter which subset it is added to, then it should have zero value. \textbf{b. Symmetry:} If two features contribute equally to the model, then their effect must be the same. \textbf{c. Linearity:} The effect a feature has on the sum of two
function is the effect it has on one function, plus the effect it has on the other~\cite{Young1985MonotonicSO}.
  \begin{equation}
      \begin{split}
          \text{If } V_1(\mathcal{S} \cup \{k\}) - V_1(\mathcal{S}) \geq V_2(\mathcal{S} \cup \{k\}) - V_2(\mathcal{S}),\\
  \text{ for all } \mathcal{S} \subseteq \mathcal{D} \setminus \{k\}
  \Rightarrow \mathcal{X}_k(V_1) \geq \mathcal{X}_k(V_2).
      \end{split}
  \end{equation} 
\noindent\newtheorem{theorem}{Theorem}
\begin{theorem}[Shapley Mode Value]
Any Mode valuation $\mathcal{X}_k(\mathcal{D}, \mathcal{Z}, V)$ satisfying Axioms 1 and 2 must have the form:
\begin{equation}\label{Eq:ShapleyValue}
\hspace{-1mm}\mathcal{X}_k = \sum_{\mathcal{S} \subseteq \mathcal{D} \setminus \{k\}} \frac{|\mathcal{S}|!(|\mathcal{D}|-|\mathcal{S}|-1)!}{|\mathcal{D}|!}\big[V(\mathcal{S} \cup \{k\}) - V(\mathcal{S})\big],
\end{equation}
where $\mathcal{X}_k$ is called the ``Shapley Mode value" of mode $k$. If and only if we calculate Eq.~\ref{Eq:ShapleyValue} for valuation, then Axioms 1 and 2 are satisfied.
\end{theorem}

\textbf{Proof.} The expression of Eq.~\ref{Eq:ShapleyValue} is the same as the
Shapley value is defined in GT~\cite{shapley:book1952}. This motivates calling
Eq.~\ref{Eq:ShapleyValue} the data Shapley Mode Value. The mathematical proof comes from the fact that the original Shapley value is the unique solution to a specific type of problem, and we can transform the mode valuation problem into the same mathematical framework. To understand this connection, consider a cooperative game in which multiple players collaborate to achieve a shared goal. In GT, you have a certain number of players, and there's a function that tells you what reward or score any group of these players can achieve if they collaborate. The key question becomes: if different groups of players generate different levels of success, how do you fairly distribute the total reward among all the individual players? 

Shapley developed a mathematical solution to determine each player's fair share of the reward. Here, fairness is codified by properties that are mathematically
equivalent to the axioms that we listed. We can view mode valuation as a cooperative game: each mode acts like a player in the game. When you combine different subsets of modes you get different levels of prediction performance, just like different groups of players achieve different scores. The Shapley mode value then works precisely like the payment system in the cooperative game, determining how much credit each mode deserves based on its contribution to the overall model performance.
\subsection{Approximating Shapley Mode Value based on Monte Carlo}
Computing the exact Shapley value involves $2^Z$ subsets, which is intractable. Inspired by~\cite{Data-Shapleyghorbani19c}, rewriting Eq.~\ref{Eq:ShapleyValue} by letting $\mathbf{U}$ be a uniform permutation of $\mathcal{D}$:
\begin{equation}
\mathcal{X}_k = \mathbb{E}_\mathbf{U} \left[ V(\mathcal{S}^{\mathbf{U}}_k \cup \{k\}) - V(\mathcal{S}^{\mathbf{U}}_k) \right],
\end{equation}
\noindent where $\mathcal{S}^{\mathbf{U}}_k$ is the set of elements before $k$ in permutation $\mathbf{U}$. The Monte Carlo algorithm involves Sampling permutations, scanning through each permutation to compute the marginal contribution, and averaging the contributions across permutations. This is an unbiased estimator of the Shapley mode value. This reduces the complexity from exponential to polynomial, specifically $O(P \cdot Z)$, where $P$ is the number of Monte Carlo permutations and $Z$ is the number of modes. In practice, $P \ll 2^Z$ due to the convergence criterion in Eq.~(\ref{eq: convergence}). Since $V(\mathcal{S})$ is computed on a finite test set, it contains noise. Moreover, as $|\mathcal{S}|$ increases, the marginal contribution of a single mode becomes negligible. So, if $|V(D) - V(\mathcal{S})|$ is less than a performance tolerance (based on bootstrap variation), we stop computing further contributions. Performance tolerance is the mean relative absolute error between iterations:
\begin{equation}
\frac{1}{Z}\sum_{k=1}^{Z} \frac{|\mathcal{X}_{k}^{t} - \mathcal{X}_{k}^{t-100}|}{|\mathcal{X}_{k}^{t}|} < 0.01.\label{eq: convergence}
\end{equation}
Convergence is reached when the average change in Shapley values ($\mathcal{X}_{k}$) over 100 iterations is less than 1\%. Algorithm~\ref{Monte Carlo Shapley Value} provides pseudo-code for the SMV steps. 
\begin{algorithm}[t]
\caption{Shapley Mode Value Approximation}\label{Monte Carlo Shapley Value}
\begin{algorithmic}
\Function{Monte Carlo Shapley} {$\mathcal{D}, \mathcal{Z}, V, \epsilon_3$}
    \State Initialize $\mathcal{X}_k = 0$ for $k = 1, \dots, Z$, and $t=0$
    \While {Convergence criteria are not met} 
        \State $t \gets t+1$
        \State $\mathbf{U}^t \gets \text{Random Permutation of data points}$
        \State $v_0^t \gets V(\emptyset,\mathcal{Z})$
        \For{$j = 1$ to $Z$}
            \If{$|V(\mathcal{D}) - v_{j-1}^t| < \epsilon_3$}
                \State $v_j^t \gets v_{j-1}^t$
            \Else
                \State $v_j^t \gets V(\{\mathbf{U}^t[1], \dots, \mathbf{U}^t[j]\},\mathcal{Z})$
            \EndIf
            \State $\mathcal{X}_{k} \gets \frac{t-1}{t}\mathcal{X}_{k} + \frac{1}{t}(v_j^t - v_{j-1}^t)$
        \EndFor
    \EndWhile
    \State \textbf{Sort} modes such that $\mathcal{X}_{1} \ge \mathcal{X}_{2} \ge \dots \ge \mathcal{X}_{Z}$
    \State \textbf{Return} top $K$ modes $\mathcal{M}^A_K= \{m_l^{\mathcal{X}_l}\}_{l=1}^K, \quad K\leq Z$
\EndFunction
\end{algorithmic}
\end{algorithm}

\section{Experiments \& Performance Evaluation}\label{Exp}

This section describes the experimental setup and comprehensive performance evaluation.

\subsection{Datasets \& Neural Networks}
\begin{figure}
    \centering
    \includegraphics[width=\linewidth]{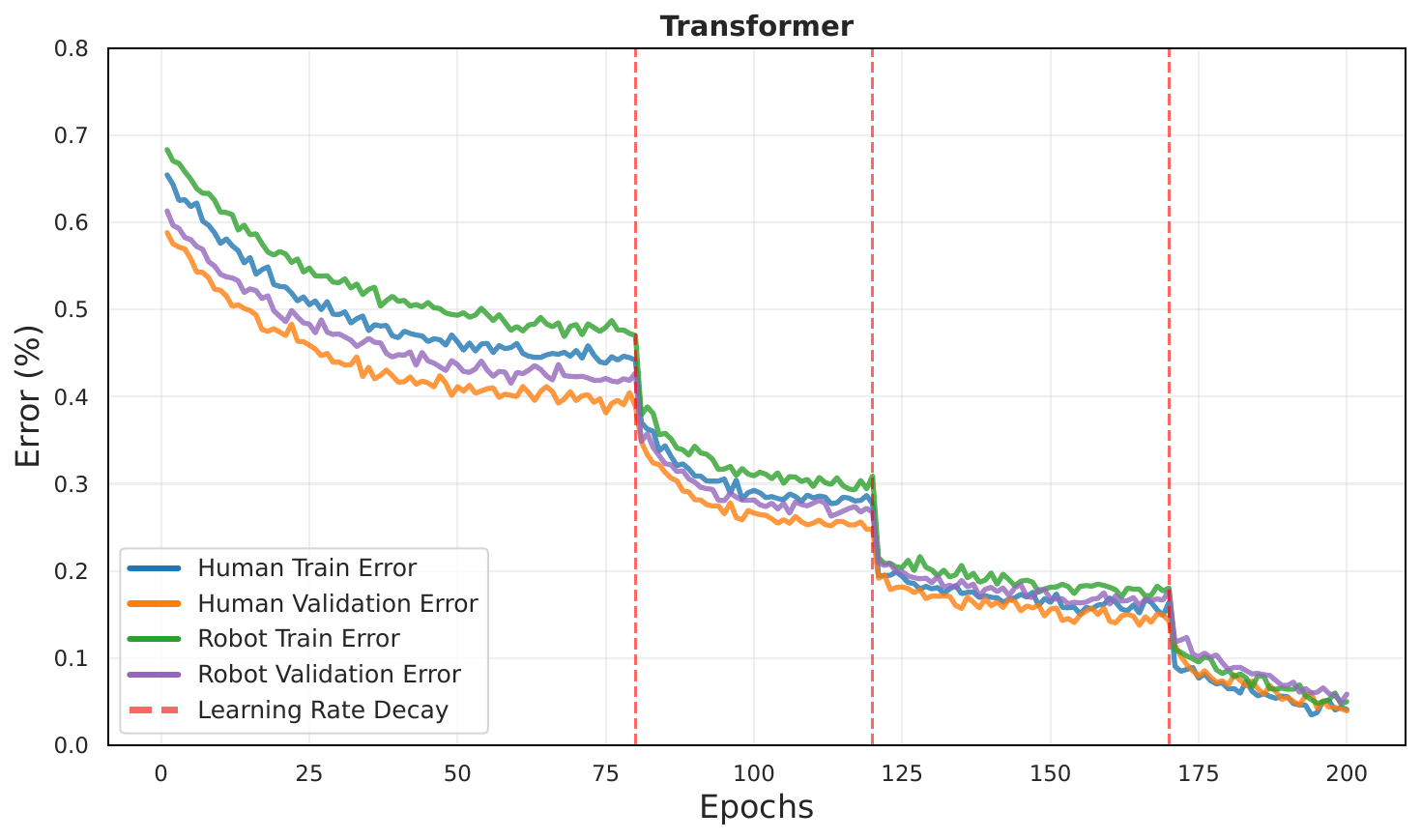}
    \caption{Error evaluation of Transformer in training phase $W$=1.}
    \label{fig: Evaluation}
\end{figure}
\begin{figure*}
    \centering
    \includegraphics[width=\textwidth]{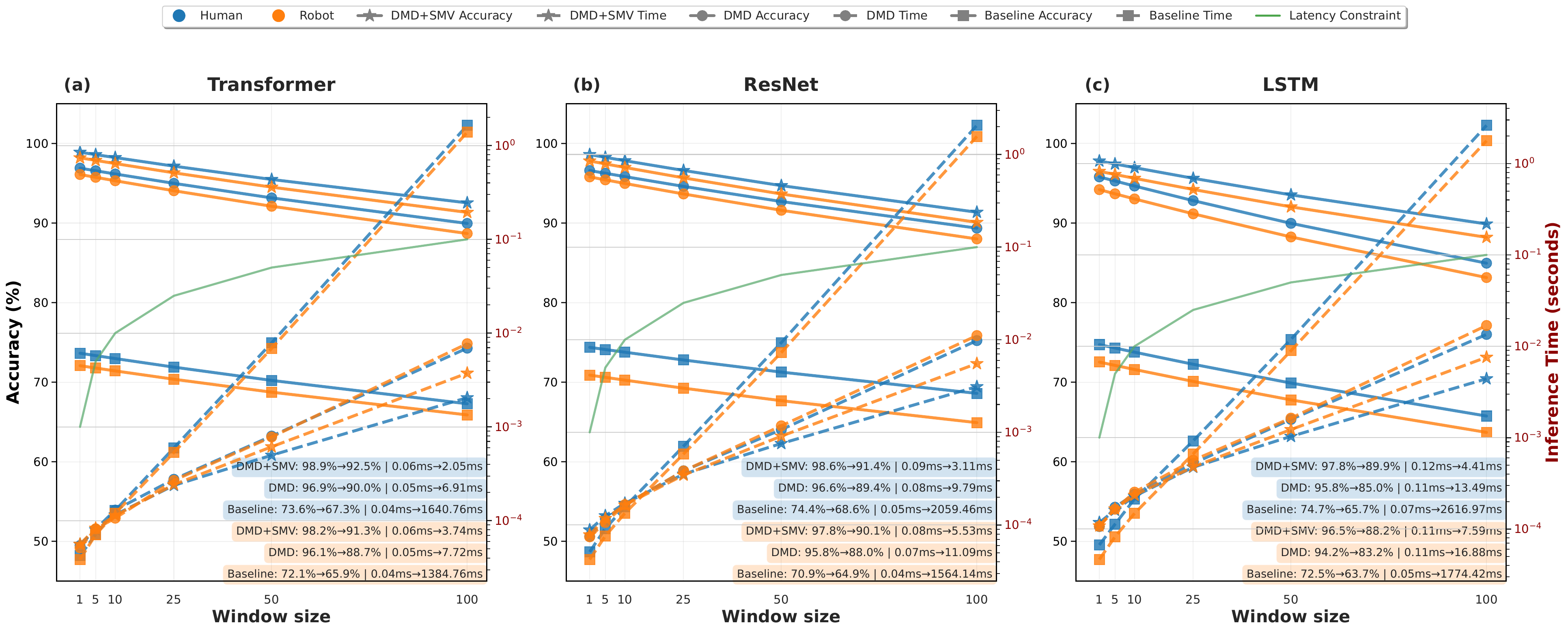}
    \caption{Accuracy and inference time evaluation of DMD and DMD+SMV, and baseline prediction for three NN architectures. The changes of accuracy and inference time from W=1 to 100 are provided for each method. Solid lines indicate prediction accuracy (\%) based on the left y-axis, while dashed lines represent inference time (seconds) based on the right y-axis. A green line, aligned with the right y-axis, is the latency constraint.}
    \label{fig: Combine}
\end{figure*}

\textbf{Datasets:} We utilized real-world haptic data traces in the experiments. The dataset captures kinaesthetic interactions recorded using a Novint Falcon haptic device within a Chai3D virtual environment. The dataset provides detailed records of 3D position, velocity, and force measurements~\cite{rodriguez_guevara_2025_14924062}. Model performance and inference time were evaluated on the ``Tap-and-Hold" dataset, which was treated as an unseen test set.

\textbf{Neural Networks:} We employed three NN architectures pre-trained on 10 diverse TI datasets with two NVIDIA RTX A6000 GPUs. \textbf{1. Transformers~\cite{VaswaniNIPS2017_3f5ee243}:} 
The attention mechanism is incorporated into the last layer. \textbf{2. ResNet-32~\cite{He_2016_CVPR}:} The network weights were initialized using He initialization~\cite{He_2015_ICCV}. \textbf{3. LSTM~\cite{LSTM6795963}:}
LSTM consists of two stacked LSTM layers with 128 units each, followed by a dense output layer with a linear activation function.

\subsection{Training \& Inference Phase}
The Adam optimizer was employed for training, with default exponential decay rates for the first and second moment estimates set to 0.9 and 0.999, respectively, as recommended in the original formulation~\cite{adamkingma2017adammethodstochasticoptimization}. The initial learning rate was set to 0.001.
All three NN architectures were trained for a maximum of 200 epochs. To encourage stable convergence and prevent overfitting, a learning rate decay schedule was used, reducing the learning rate by a factor of 0.005 at epochs 80, 120, and 170. To regularize training and improve generalization, Dropout~\cite{Dropout10.5555/2627435.2670313} was applied to all models. A dropout rate of P-drop = 0.1 was used for the ResNet-32 and Transformer architectures, while a slightly higher rate of P-drop = 0.2 was used for the LSTM, due to its higher tendency to overfit on temporal data. All results are averaged over 10 independent runs, with training using 70\% of the data and testing using 20\%, and 10\% of the data reserved as a validation set. Fig.~\ref{fig: Evaluation} illustrates the error trend per epoch during training of the Transformer architecture, providing insights into the model's convergence behavior. We do not show the results of RestNet-32 and LSTM due to space limitations, as they follow the same trend with different start and end points.

In our experiment, we evaluate prediction accuracy and inference time across different prediction window sizes ($W \in \{1, 5, 10, 25, 50, 100\}$,) for NN architectures, each tasked with predicting the next 100 samples. We compared DMD and DMD+SMV with the baseline method, which is signal prediction from raw data without any decomposition. Note, VMD, SVMD, and VME are designed for continuous signals; therefore, a direct comparison with them is not possible. In our framework, the actual samples are used to update the signal modes for subsequent predictions. Therefore, as long as these samples arrive before the next mode update (e.g., every 100 samples), the prediction process remains effective. Namely, packet retransmission within 100 ms, which can be easily realised in communication systems.

\subsection{Accuracy Comparison} 
\begin{figure}[tbp]
    \centering
    \includegraphics[width=\linewidth]{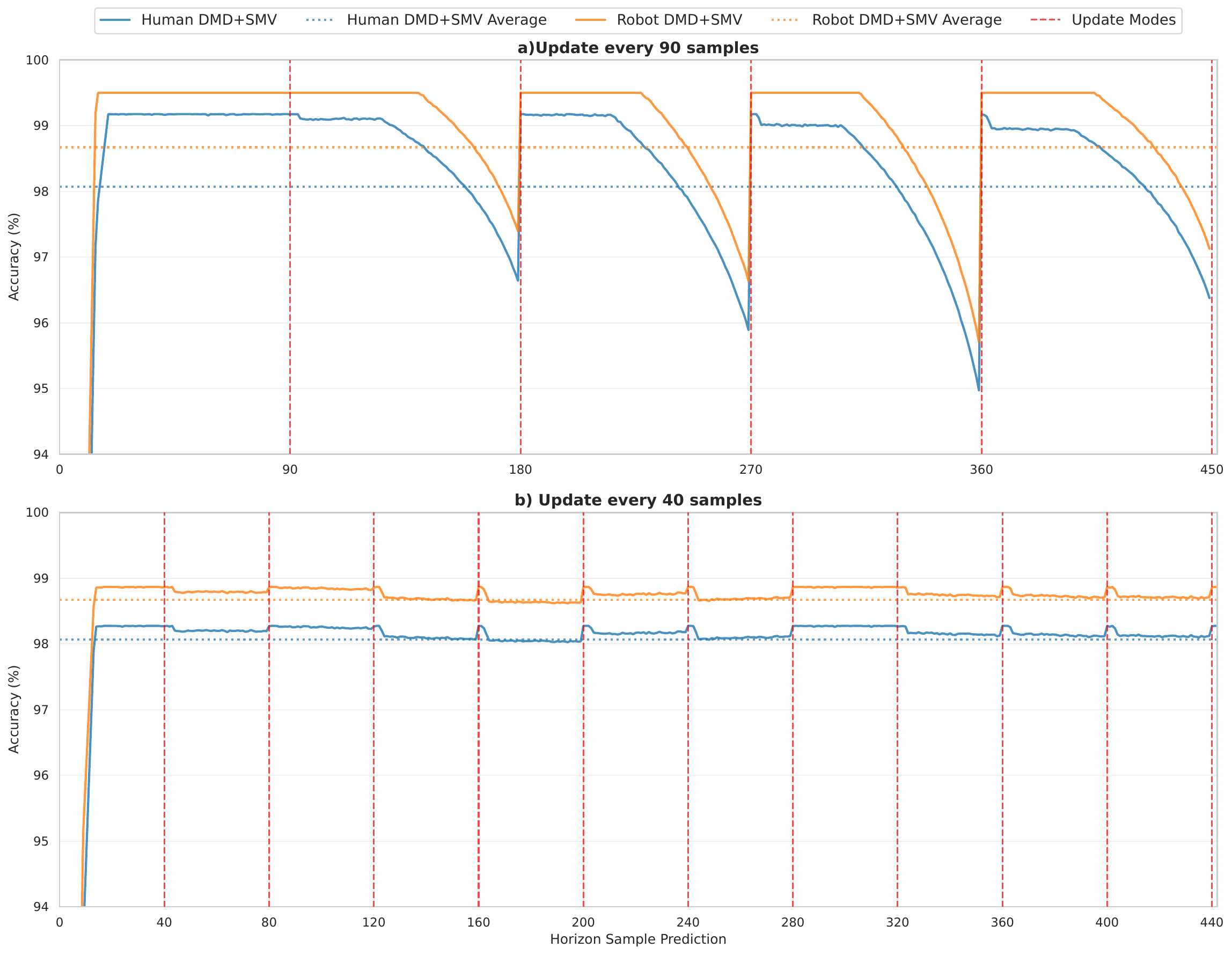}
    \caption{Accuracy evaluation of Transformer architectures for sliding  $W$=5 in the inference phase.}
    \label{fig: sliding1}
\end{figure}

Fig.~\ref{fig: Combine}~(a) demonstrates that the accuracy of the Transformer model, using both DMD and DMD+SMV, outperforms ResNet-32, Fig.~\ref{fig: Combine}~(b), and LSTM, Fig.~\ref{fig: Combine}~(c), across varying prediction window sizes ($W$). Notably, the Transformer with DMD+SMV at $W = 1$ achieves the best accuracy of approximately 98.9\% on the human side and 98.2\% on the robot side. In this experiment, the modes are updated upon completion of 100 samples. Note that increasing the modes update frequency can further improve accuracy, as we will discuss in Fig.~\ref{fig: sliding1}. This superior performance can be attributed to the Transformer’s attention mechanism and tokenization strategy. In addition, the smaller window sizes enable the model to focus on more immediate and relevant temporal features, thereby enhancing its prediction accuracy. 

\begin{figure}[tbp]
    \centering
\includegraphics[width=\linewidth]{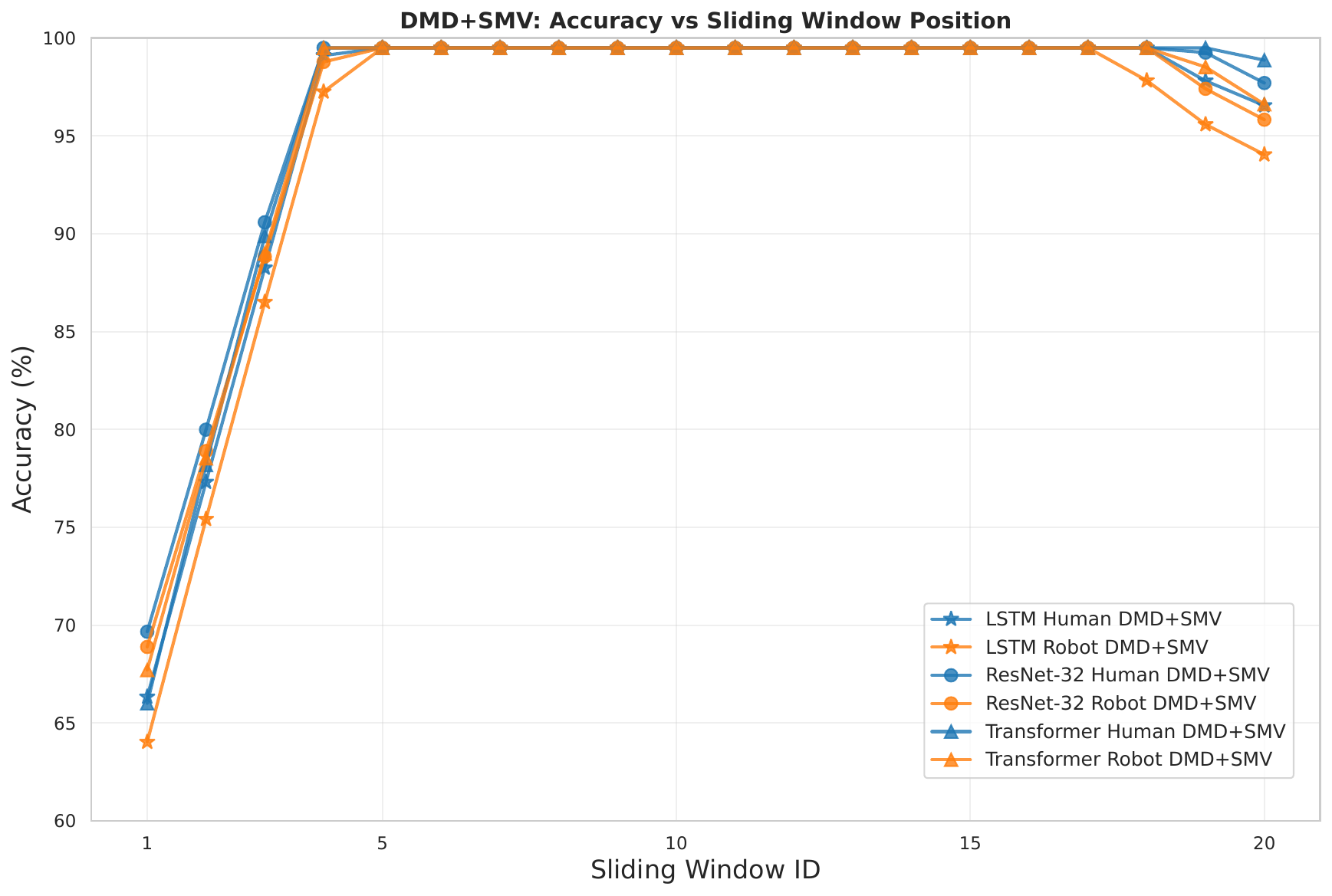}
    \caption{Inference Accuracy changes over sliding windows with $W=5$ for predicting next 100-samples.}
    \label{fig: sliding}
\end{figure}

Each point in Fig.~\ref{fig: Combine} represents the average accuracy across of all window slides for a specific $W$. For example, when $W = 5$, the point reflects the average accuracy over 20 window slides for the next 100-sample prediction. To examine this in greater detail, Fig.~\ref{fig: sliding} illustrates the accuracy changes across each sliding window for the 100-sample prediction with $W = 5$. The results show that the lowest accuracy occurs during the first four sliding windows (corresponding to the initial 20 samples). This reduced accuracy is attributed to the model’s limited ability to learn the interaction dynamics between the human and robot at such an early stage. However, this scenario would only arise if a communication outage occurred within the first 20 ms of the interaction, a situation that is highly unlikely in practical settings. Furthermore, this initial 20 ms window represents a negligible fraction of typical teleoperation tasks, and the system can rely on simple extrapolation or hold the last sample strategies during this brief initialization without compromising safety. After this initial period, the accuracy stabilizes around 99.53\%, but after about 17 prediction windows, the accuracy starts to decline due to the obsolescence of the learned modes over time.

An ablation study was conducted in which the mode update interval was reduced to mitigate the issue of accuracy drop at the end of the period in Fig.~\ref{fig: sliding}. Fig.~\ref{fig: sliding1} (a) shows the accuracy when the update interval was set to every 90 samples, which yields an average of 98.07\% and 98.67\% for the human and robot side, respectively. This strategy effectively reduces the accuracy degradation observed at the end of each sub-horizon, shows improvement compared to 100 modes update, which achieves accuracy 97.6\% and 97.87\% for human and robot, respectively. Though it incurs a 35\% increase in training time. The results show that applying DMD+SMV to the Transformer model during inference leads to more stable and reliable accuracy throughout the sample prediction horizon. To further address the performance drop near the end of each sub-horizon, modes were updated more frequently, every 40 samples as shown in Fig.~\ref{fig: sliding1} (b), which proved essential for maintaining accuracy which yielding an average of 98.22\% and 98.91\% for the human and robot side, respectively.  However, this improvement comes with a notable trade-off: a 144\% increase in training time. Since training is done offline, an update interval of 40 samples is acceptable for real-world implementation. 

\begin{figure}[tbp]
    \centering
    \includegraphics[width=\linewidth]{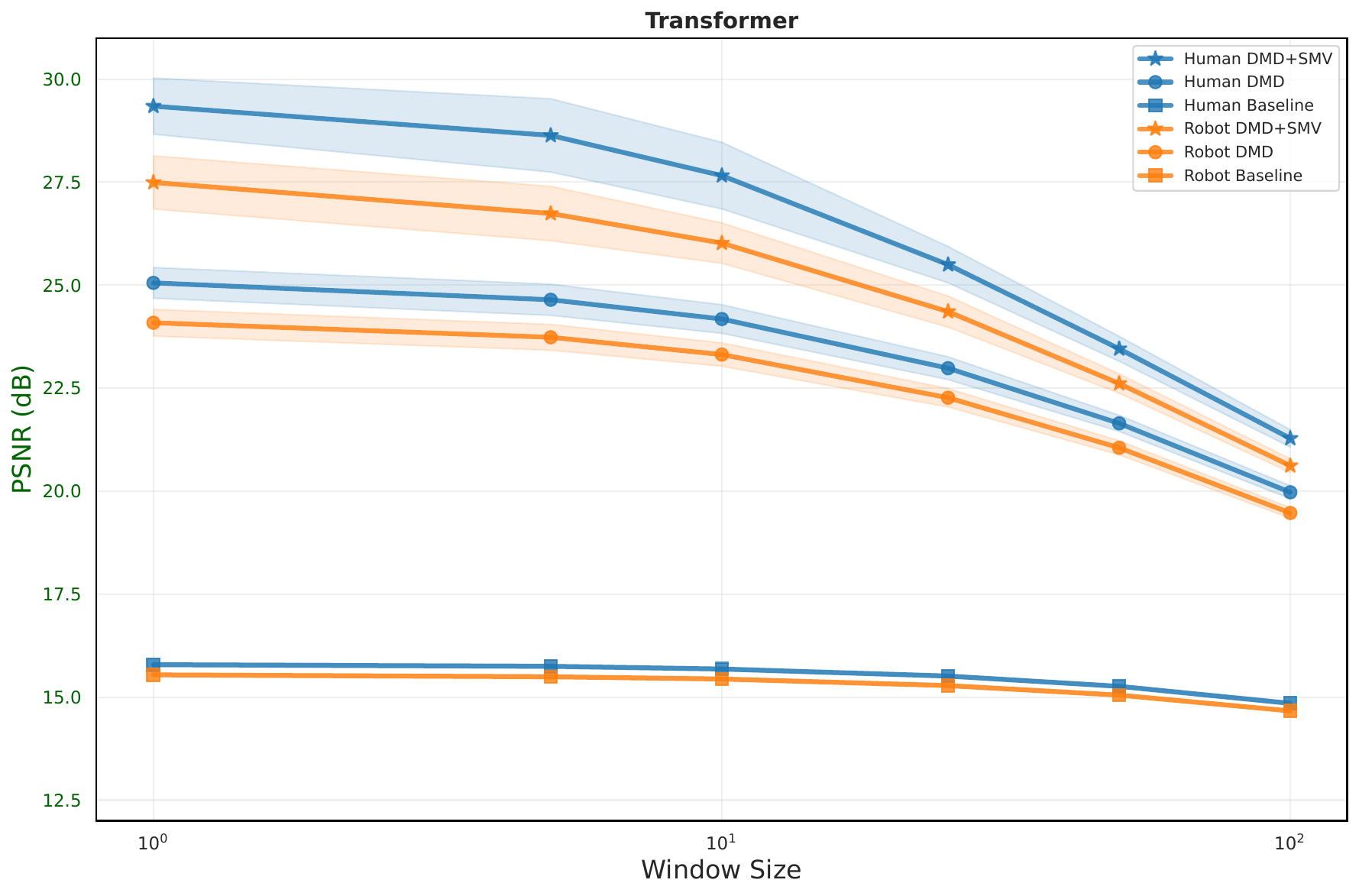}
    \caption{Transformer architecture: PSNR comparison over different window sizes. Shaded colors are the standard deviation.}
    \label{fig: PSNR}
\end{figure}
In Fig.~\ref{fig: PSNR}, we present the Peak Signal-to-Noise Ratio (PSNR) for Transformer. Notably, the overall PSNR trends are consistent across NNs (results for the ResNet and LSTM provided in the code repository). The Transformer architecture with DMD+SMV achieves approximately 29.5 $dB$ and 27.5 $dB$ at $W=1$ for human and robot, respectively. As the prediction horizon extends to $W=100$, PSNR drops by 8.5 $dB$ and 7 $dB$ for human and robot, respectively, demonstrating the challenge of maintaining accuracy over longer horizons. It can be seen that DMD provides substantial improvements over the baseline methods, achieving ~9-10 $dB$ gains for both human and robot data across all prediction horizons compared to their respective baselines (15 $dB$). The DMD+SMV enhancement delivers an additional 4-5 $dB$ improvement over standard DMD.

Each point in Fig.~\ref{fig: sliding} represents the average accuracy across 9 outputs corresponding to the three spatial dimensions (X, Y, Z) of the three features: velocity, force, and position. To provide a more detailed view of Fig.~\ref{fig: sliding}, the average of each three features in three dimensions is provided in Table~\ref{Tab: Tap and Hold} at window sizes of $W=1$ and $W=5$.
On average, features predicted using the Transformer combined with DMD+SMV achieve higher accuracy. While both human and robot predictions benefit from this approach, the robot-side generally yields higher accuracy across most features. Overall, across all experiments, the DMD+SMV method consistently outperforms both the standard DMD approach and the baseline. This suggests that the SMV arms the DMD with more robust mode selection.

\begin{table*}[tbp]
\centering
\caption{Accuracy results averaged over 10 runs for window sizes $W\in\{1, 5\}$. Results are compared between the Baseline, DMD, and DMD+SMV methods across LSTM, ResNet-32, and Transformer on the "Tap and Hold" dataset.}\label{Tab: Tap and Hold}
\resizebox{\textwidth}{!}{
\begin{tabular}{c|l|cccccc|cccccc}
\multicolumn{2}{c}{\textbf{Window size}} & \multicolumn{6}{c}{\textbf{$W=1$}} & \multicolumn{6}{c}{\textbf{$W=5$}} \\
\cline{1-14}
\multirow{2}{*}{NN} & \multirow{2}{*}{Features} & \multicolumn{2}{c}{\textbf{Baseline}} & \multicolumn{2}{c}{\textbf{DMD}} & \multicolumn{2}{c|}{\textbf{DMD+SMV}} & \multicolumn{2}{c}{\textbf{Baseline}} & \multicolumn{2}{c}{\textbf{DMD}} & \multicolumn{2}{c}{\textbf{DMD+SMV}} \\
\cline{3-14}
 & & Human & Robot & Human & Robot & Human & Robot & Human & Robot & Human & Robot & Human & Robot \\
\hline
\multirow{9}{*}{\rotatebox{90}{\textbf{LSTM}}}
& Position X & 74.91 ± 2.18 & 72.73 ± 1.94 & 96.05 ± 2.84 & 94.37 ± 1.62 & 98.14 ± 1.25 & 96.82 ± 2.46 & 73.18 ± 2.04 & 71.25 ± 1.72 & 94.41 ± 2.19 & 92.84 ± 1.43 & 96.54 ± 0.98 & 95.27 ± 2.36 \\
& Position Y & 75.26 ± 1.75 & 72.18 ± 2.39 & 95.49 ± 1.37 & 93.94 ± 2.18 & 97.53 ± 1.64 & 96.38 ± 1.57 & 75.06 ± 1.63 & 73.38 ± 2.05 & 96.13 ± 1.15 & 94.51 ± 2.48 & 98.54 ± 1.31 & 97.05 ± 1.27 \\
& Position Z & 74.17 ± 2.51 & 72.89 ± 1.46 & 96.32 ± 2.17 & 94.75 ± 1.84 & 98.26 ± 1.39 & 96.94 ± 2.28 & 72.48 ± 2.92 & 69.85 ± 1.58 & 93.47 ± 2.64 & 91.94 ± 1.26 & 95.26 ± 1.48 & 94.32 ± 2.06 \\
& Velocity X & 75.08 ± 1.93 & 72.64 ± 2.17 & 95.74 ± 1.85 & 94.29 ± 2.06 & 97.69 ± 1.47 & 96.61 ± 1.73 & 75.42 ± 1.77 & 72.88 ± 2.52 & 96.05 ± 1.43 & 94.85 ± 2.20 & 98.61 ± 1.25 & 96.93 ± 1.86 \\
& Velocity Y & 74.56 ± 2.04 & 72.31 ± 2.84 & 95.87 ± 1.76 & 94.18 ± 2.57 & 97.84 ± 1.28 & 96.47 ± 1.91 & 73.15 ± 1.94 & 71.29 ± 2.36 & 94.47 ± 1.58 & 92.51 ± 2.08 & 96.38 ± 1.03 & 95.26 ± 1.42 \\
& Velocity Z & 74.39 ± 2.73 & 72.07 ± 1.65 & 96.18 ± 2.29 & 94.52 ± 1.38 & 98.07 ± 1.16 & 96.73 ± 2.14 & 76.51 ± 2.58 & 74.04 ± 1.77 & 97.16 ± 2.41 & 95.81 ± 1.54 & 99.15 ± 1.12 & 98.38 ± 1.95 \\
& Force X & 75.31 ± 1.82 & 72.95 ± 2.68 & 95.62 ± 1.94 & 94.06 ± 2.31 & 97.58 ± 1.53 & 96.29 ± 1.86 & 77.19 ± 1.54 & 75.26 ± 2.72 & 96.42 ± 1.28 & 94.53 ± 2.47 & 98.22 ± 1.04 & 97.38 ± 1.48 \\
& Force Y & 74.83 ± 2.26 & 72.45 ± 2.52 & 96.04 ± 2.37 & 94.38 ± 2.19 & 97.91 ± 1.42 & 96.55 ± 1.24 & 75.37 ± 2.39 & 73.05 ± 2.58 & 96.18 ± 2.05 & 94.82 ± 2.27 & 98.59 ± 1.07 & 96.97 ± 1.03 \\
& Force Z & 74.92 ± 2.61 & 72.76 ± 1.79 & 95.53 ± 2.08 & 94.11 ± 1.52 & 97.73 ± 1.61 & 96.84 ± 1.97 & 71.48 ± 2.72 & 69.24 ± 1.85 & 92.06 ± 2.27 & 90.52 ± 1.46 & 94.63 ± 1.18 & 93.27 ± 1.30 \\
\hline
\multirow{9}{*}{\rotatebox{90}{\textbf{ResNet}}}
& Position X & 74.73 ± 2.41 & 71.19 ± 1.87 & 96.84 ± 1.73 & 96.04 ± 0.95 & 98.97 ± 0.34 & 98.15 ± 0.18 & 72.98 ± 2.37 & 69.71 ± 1.58 & 95.47 ± 1.50 & 94.28 ± 0.82 & 97.32 ± 0.29 & 96.69 ± 0.15 \\
& Position Y & 74.05 ± 1.84 & 70.87 ± 2.47 & 96.29 ± 1.06 & 95.72 ± 1.59 & 98.31 ± 0.95 & 97.64 ± 1.12 & 75.24 ± 1.74 & 71.47 ± 2.18 & 97.08 ± 0.92 & 95.69 ± 1.26 & 99.13 ± 0.82 & 98.27 ± 0.87 \\
& Position Z & 74.96 ± 2.35 & 70.62 ± 1.96 & 96.71 ± 1.48 & 95.47 ± 1.13 & 98.65 ± 1.79 & 97.93 ± 0.67 & 72.38 ± 2.16 & 68.43 ± 1.85 & 94.52 ± 1.29 & 92.15 ± 0.98 & 96.04 ± 1.33 & 95.69 ± 0.49 \\
& Velocity X & 74.29 ± 1.59 & 71.08 ± 3.14 & 96.51 ± 0.52 & 95.93 ± 2.35 & 98.73 ± 1.07 & 97.84 ± 1.78 & 73.83 ± 1.28 & 70.94 ± 2.86 & 96.48 ± 0.38 & 94.16 ± 2.06 & 98.05 ± 0.83 & 97.62 ± 1.44 \\
& Velocity Y & 74.51 ± 1.64 & 70.39 ± 2.58 & 96.18 ± 0.79 & 95.26 ± 1.71 & 98.29 ± 0.35 & 97.49 ± 0.73 & 76.32 ± 1.42 & 72.41 ± 2.16 & 98.03 ± 0.69 & 96.78 ± 1.25 & 98.26 ± 0.27 & 97.44 ± 0.60 \\
& Velocity Z & 74.17 ± 1.83 & 71.47 ± 1.92 & 96.73 ± 1.95 & 95.64 ± 1.05 & 98.52 ± 1.74 & 97.86 ± 0.54 & 74.93 ± 1.54 & 71.85 ± 1.76 & 97.42 ± 1.69 & 95.21 ± 0.92 & 98.47 ± 1.43 & 98.17 ± 0.38 \\
& Force X & 74.64 ± 2.76 & 70.83 ± 1.29 & 96.19 ± 1.84 & 95.18 ± 0.41 & 98.41 ± 1.86 & 97.52 ± 1.14 & 71.38 ± 2.50 & 67.26 ± 1.04 & 93.07 ± 1.53 & 91.69 ± 0.28 & 95.48 ± 1.43 & 94.20 ± 0.87 \\
& Force Y & 74.25 ± 3.58 & 70.74 ± 1.61 & 96.84 ± 2.67 & 95.91 ± 0.68 & \cellcolor{green!25}98.94 ± 0.89 & 98.03 ± 1.73 & 74.26 ± 3.16 & 70.84 ± 1.26 & 96.53 ± 2.25 & 94.17 ± 0.46 & 98.02 ± 0.68 & 97.73 ± 1.36 \\
& Force Z & 74.52 ± 2.17 & 70.95 ± 2.48 & 96.37 ± 1.12 & 95.06 ± 1.64 & 98.26 ± 2.05 & 97.41 ± 0.81 & 72.85 ± 1.93 & 69.73 ± 2.16 & 95.48 ± 0.84 & 93.16 ± 1.29 & 97.51 ± 1.63 & 96.22 ± 0.58 \\
\hline
\multirow{9}{*}{\rotatebox{90}{\textbf{Transformer}}}
& Position X & 73.82 ± 1.68 & 72.31 ± 4.16 & 97.14 ± 1.72 & 96.35 ± 2.18 & \cellcolor{green!25}99.17 ± 1.06 & \cellcolor{cyan!30}98.47 ± 0.62 & 72.48 ± 1.02 & 70.63 ± 3.79 & 95.73 ± 1.41 & 94.92 ± 1.83 & 97.45 ± 0.88 & 96.72 ± 0.48 \\
& Position Y & 73.29 ± 3.17 & 72.14 ± 1.72 & 97.53 ± 1.14 & 96.08 ± 1.68 & 99.34 ± 1.85 & \cellcolor{cyan!30}98.76 ± 0.02 & 74.18 ± 2.86 & 72.95 ± 1.55 & 97.83 ± 0.87 & 95.58 ± 1.43 & \cellcolor{green!25}99.75 ± 1.46 & 98.44 ± 0.01 \\
& Position Z & 73.18 ± 2.74 & 71.93 ± 4.28 & 97.28 ± 1.66 & 96.24 ± 2.00 & \cellcolor{green!25}99.06 ± 0.04 & \cellcolor{cyan!30}98.52 ± 0.00 & 73.51 ± 2.46 & 71.63 ± 3.85 & 96.32 ± 1.48 & 94.98 ± 1.84 & 98.74 ± 0.03 & 97.63 ± 0.00 \\
& Velocity X & 73.61 ± 3.85 & 71.75 ± 4.37 & 97.09 ± 2.86 & 95.84 ± 2.10 & \cellcolor{green!25}98.93 ± 0.63 & \cellcolor{cyan!30}98.24 ± 2.19 & 71.19 ± 3.19 & 69.94 ± 3.93 & 94.82 ± 2.46 & 92.51 ± 1.87 & 96.21 ± 0.48 & 95.73 ± 1.86 \\
& Velocity Y & 73.18 ± 2.97 & 72.09 ± 4.05 & 97.42 ± 1.15 & 96.33 ± 0.85 & \cellcolor{green!25}99.68 ± 1.29 & \cellcolor{cyan!30}98.76 ± 0.61 & 74.51 ± 2.68 & 72.62 ± 3.63 & 97.33 ± 0.86 & 95.98 ± 0.68 & 99.47 ± 1.03 & 98.63 ± 0.48 \\
& Velocity Z & 73.94 ± 1.73 & 72.05 ± 1.07 & 96.87 ± 0.58 & 95.63 ± 2.57 & \cellcolor{green!25}98.73 ± 1.23 & \cellcolor{cyan!30}98.19 ± 0.62 & 72.16 ± 1.14 & 70.94 ± 0.86 & 95.82 ± 0.46 & 93.51 ± 2.18 & 97.48 ± 0.97 & 96.73 ± 0.39 \\
& Force X & 73.21 ± 4.96 & 71.84 ± 0.89 & 97.16 ± 2.52 & 96.05 ± 3.14 & \cellcolor{green!25}98.79 ± 1.12 & \cellcolor{cyan!30}98.34 ± 0.64 & 71.52 ± 4.48 & 69.63 ± 0.68 & 94.33 ± 2.28 & 92.92 ± 2.64 & 96.75 ± 0.86 & 95.63 ± 0.44 \\
& Force Y & 73.08 ± 4.17 & 71.65 ± 4.28 & 97.34 ± 2.04 & 95.92 ± 2.99 & 98.61 ± 2.11 & \cellcolor{cyan!30}98.19 ± 1.58 & 73.19 ± 3.66 & 71.94 ± 3.85 & 96.82 ± 1.67 & 94.51 ± 1.84 & 98.15 ± 1.94 & 97.63 ± 1.28 \\
& Force Z & 73.87 ± 2.74 & 71.26 ± 5.28 & 97.19 ± 1.76 & 96.03 ± 2.43 & \cellcolor{green!25}99.02 ± 0.95 & \cellcolor{cyan!30}98.37 ± 1.18 & 75.51 ± 2.46 & 73.63 ± 4.86 & 96.82 ± 1.48 & 94.98 ± 1.84 & 98.60 ± 0.75 & 97.88 ± 0.93 \\
\bottomrule
\end{tabular}}
\end{table*}
\begin{table}[tbp]
\centering
\caption{Method-wise speed comparison ($W=100$).}\label{Table: method-wise}
\begin{tabular}{lcccccc}
\toprule
\multirow{2}{*}{DMD+SMV} & \multicolumn{3}{c}{\textbf{Human Side}} & \multicolumn{3}{c}{\textbf{Robot Side}} \\
\cmidrule(lr){2-4} \cmidrule(lr){5-7}
& vs DMD & vs Baseline & & vs DMD & vs Baseline & \\
\midrule
Transformer & 3.45$\times$ & 820.4$\times$ & & 2.08$\times$ & 374.3$\times$ & \\
ResNet & 3.16$\times$ & 664.4$\times$ & & 2.02$\times$ & 284.4$\times$ & \\
LSTM & 3.07$\times$ & 594.3$\times$ & & 2.22$\times$ & 233.5$\times$ & \\
\bottomrule
\end{tabular}
\end{table}
\begin{table}[t]
\centering
\setlength{\tabcolsep}{1mm}
\caption{Inference time ($ms$) ($W=100$). P, V, and F stand for Position, Velocity, and Force, respectively.}
\label{tab: inference_times}
\begin{tabular}{lccccccccc}
\toprule
\multirow{2}{*}{Architecture} & \multicolumn{4}{c}{\textbf{DMD}} & \multicolumn{5}{c}{\textbf{DMD+SMV}} \\
\cmidrule(lr){2-5} \cmidrule(lr){6-10}
 & P & V & F & Avg & P & V& F & Avg & $\uparrow$ \\
\midrule
LSTM & 14.1 & 13.46 & 18.04 & 15.2 & 5.52 & 5.28 & 7.08 & 6.00 & 60.5\% \\
ResNet-32 & 9.82 & 9.20 & 12.33 & 10.45 & 3.75 & 3.88 & 5.27 & 4.30 & 58.9\% \\
Transformer & 6.87 & 6.42 & 8.61 & \cellcolor{green!25}7.30 & 2.61 & 2.50 & 3.35 & \cellcolor{green!25}2.85 & \cellcolor{green!25}61\% \\
\bottomrule
\end{tabular}
\end{table}
\begin{table}[t]
\centering
\caption{FLOPs (Floating Point Operations) and FLOPS (Floating Point Operations per Second) compared between the DMD and DMD+SMV methods.}\label{Tab: FLOPs}
\begin{tabular}{lcccc}
\toprule
\multirow{2}{*}{Architecture} & \multicolumn{2}{c}{\textbf{FLOPs$\downarrow$}} & \multicolumn{2}{c}{\textbf{FLOPS $\uparrow$}}
\\ 
\cmidrule(lr){2-3} \cmidrule(lr){4-5}
 &DMD & DMD+SMV & DMD & DMD+SMV\\
\midrule
LSTM & 3.4 $\times 10^7$ & $2.1\times 10^6$ & 2.24$\times 10^9$ & 0.35$\times 10^9$ \\ 
ResNet-32 & 11.2 $\times 10^7$ & $8.6\times 10^6$ & 10.72$\times 10^9$ & 2$\times 10^9$ \\ 
Transformer & 19.3 $\times 10^7$& $14.8 \times 10^6$ & \cellcolor{green!25}26.4$\times 10^9$ & \cellcolor{green!25}5.19$\times 10^9$ \\ 
\hline
\end{tabular}
\end{table}

\subsection{Inference Time and Complexity Comparison}
As shown in Fig.~\ref{fig: Combine}, the inference time of the Transformer model using DMD+SMV for the human side increased from 0.056 ms for a 1-sample prediction to 2~ms for a 100-sample prediction. In addition, method-wise comparison in Table~\ref{Table: method-wise} demonstrates that SMV accelerates the inference dramatically compared to the DMD and Baselines. For transformer architecture, it achieves speedup by $820\times$ on the human side and $374\times$ on the robot side compared to the Baseline. The massive speedup compared to baseline occurs because the DMD+SMV model runs inference on a highly optimized, smaller feature set, whereas the Baseline must process the entire complex raw signal. In addition, DMD+SMV further accelerates the inference by $3\times$ on the human side and $2\times$ on the robot side, compared to DMD. The significant speedup of DMD+SMV compared to standard DMD is directly attributable to the mode pruning capability of the Shapley analysis. Standard DMD feeds all decomposed intrinsic modes into the predictor, including those that may represent noise or redundant information. In contrast, SMV filters out these task-irrelevant modes, reducing the dimensionality of the input feature space. Furthermore, as shown in Table~\ref{tab: inference_times}, Transformer models again demonstrate the fastest inference, averaging 2.85 ms when using DMD+SMV for both human and robot. This represents a 61\% improvement compared to DMD and is also $1.5\times$ faster than DMD+SMV with ResNet-32.

Table~\ref{Tab: FLOPs} presents a comparison of NNs based on total computational workload (FLOPs) and the rate of computation (FLOPS). The results highlight that the Transformer's attention mechanism enables a high degree of parallelism, allowing it to execute many operations simultaneously on a GPU, thereby achieving higher FLOPS. Although the FLOPs of the Transformer are large, this parallel execution substantially reduces inference time. In contrast, LSTMs are inherently sequential, which constrains their ability to exploit parallel hardware and results in slower inference. Overall, the Transformer demonstrates superior inference performance in both DMD and DMD+SMV methods, primarily due to its efficient parallelizability.

\section{Conclusion}\label{conclusion}
In this work, we presented a discrete signal decomposition framework for decomposing a signal into its fundamental components. This allows for robust prediction in TI systems. By integrating DMD for mode extraction and SMV for mode valuation, we demonstrate that DMD+SMV consistently outperforms DMD alone and baseline, as SMV provides actionable insights that identify modes contributing most to the prediction. Our framework lays the groundwork for adaptive TI systems that strike a balance between human control and semi-autonomous prediction.  As future work, we plan to integrate this framework into a real-world TI testbed to validate its performance under practical network and hardware conditions. The authors have provided public access to their code and data at https://github.com/Ali-Vahedifar/Discrete-Mode-Decomposition.git.

{
\bibliographystyle{IEEEbib}
\bibliography{refs}
}

\end{document}